\documentclass[aps,prl,twocolumn,showpacs,groupedaddress,floatfix]{revtex4}
\usepackage{amssymb,amsmath}
\usepackage{graphicx}
\usepackage{subfigure}
\usepackage[english]{babel}
\usepackage{float}
\usepackage{color}

\newcommand{\be}{\begin{equation}}
\newcommand{\ee}{\end{equation}}

\begin{document}

\title{Bulk and surface bound states in the continuum}

\author{N. A. Gallo$^{1,2}$ and M. I. Molina$^{1,3}$}

\affiliation{$^{1}$ Departamento de F\'{\i}sica,Facultad de Ciencias,Universidad de Chile,Casilla 653,Santiago,Chile\\
$^{2}$ Centro para el desarrollo de la nanociencia y la nanotecnolog\'{\i}a (CEDENNA)\\
Av. Ecuador 3493, Santiago, Chile\\
$^{3}$ Center for Optics and Photonics (CEFOP) and MSI-Nucleus on Advanced Optics, 
Casilla 4016, Concepcion, Chile}

\pacs{03.65.Nk, 03.65.Ge, 42.79.Gn}

\begin{abstract}
We examine bulk and surface bound states in the continuum (BIC) that is, square-integrable, localized modes  embedded in the linear spectral band of a discrete lattice including  interactions to first-and second nearest neighbors. We suggest an efficient method for generating such modes and the local bounded potential that supports the BIC, based on the pioneering Wigner-von Neumann concept. It is shown that the bulk and surface embedded modes are structurally stable and that they decay faster than a power law at long distances from the mode center.
\end{abstract}

\maketitle

\clearpage

\section{Introduction}
The eigenvalue structure of a general quantum system with a finite potential well consists usually of bound states and unbound continuum states, whose energies are positive if we set the asymptotic value of the potential at infinity as zero. While the bound states are localized in space and square integrable, the continuum modes are extended and non-normalizable. In 1929, however, Wigner and von Neumann suggested an exception to this picture by constructing explicitly a bound state in the continuum (BIC) that is, an eigenstate with energy above the continuum threshold but localized and square integrable\cite{wigner}. Their basic idea consisted on imposing a modulation of an otherwise sinusoidal profile with a decaying envelope that is square integrable. From that, a suitable local potential can be designed such that the proposed wavefunction is an eigenstate of this potential. Both, the potential and the wavefunction thus proposed, oscillate in space and decay as a power law. The idea lay dormant for many years, until it was retaken again by Stillinger\cite{pra10_1122} and Herrick\cite{pra11_446} who suggested that BICs might be found in certain atomic and molecular systems. Later on, they suggested the use of superlattices to construct potentials that could support BICs\cite{stillinger_physB,herrick_physB}. Subsequent experiments with semiconductor heterostructures provided the direct observation of electronic bound states above a potential well localized by Bragg reflections\cite{capasso}. A different approach for the design of potentials that can support BICs, come from the concept of resonant states in quantum mechanics. Resonant states are localized in space but with energies in the continuum, and they eventually decay, i.e., they have a finite lifetime. Under certain conditions, the interference between resonances can lead to a resonant state of zero width. In other words, the localized state decouples from the continuum becoming a BIC. One example of  this is the case of an Hydrogen atom in a uniform magnetic field, modeled as a system of coupled Coulombic channels, where interference between resonances belonging to different channels leads to the ocurrence of BICs\cite{coulombic}. More recently, BICs have been shown to occur in mesoscopic electron transport and quantum waveguides\cite{mesoscopic}, and in quantum dot systems\cite{qdot}. Here, the existence of BICs can be traced back to the destruction of the discrete-continuum decay channels by quantum interference effects. 

On the other hand, it has been admitted that the BIC phenomenon relies on interference and thus, is inherent to any wave-like system, besides quantum mechanics, such as optical systems described by the paraxial wave equation. In fact, the analogy between these two realms have gained much attention recently, and have given rise to experimental observations of many phenomena that are hard to observe in a condensed matter setting\cite{longhi}. Examples of this include dynamic localization \cite{dynamic_localization}, bloch oscillations\cite{bloch}, Zeno effect\cite{zeno} and Anderson localization\cite{anderson}. The main appeal of using optical systems is that experiments can be designed to focus on a particular aspect without the need to deal with the presence of many other effects commonly present in quantum solids, such as many-body effects. In optics it is also possible to steer and manage the propagation of excitations and to tailor the optical medium. Thus, it is no surprising that there have been also a number of recent works on BICs in classical optical systems\cite{optical_structures,molina}.

In this work we extend Wigner and von Neumann's original concept to  demonstrate that BICs can exist as bulk modes or surface modes of a discrete linear lattice, even in the presence of first-and second nearest neighbor interactions. We show explicitly that the wavefunction thus generated decays faster than a power law at long distances from the mode center, while the local potential decays as a power law. The system seems structurally stable and thus, an experimental verification of these ideas in the optical domain (waveguide arrays) looks  promising.

\section{The model}
Let us consider a linear, one-dimensional lattice in the presence of a site energy distribution $\{\epsilon_{n}\}$. In optics, this could correspond to a set of weakly
coupled optical waveguides, each of them characterized
by a propagation constant $\epsilon_{n}$ 
  and centered at $x_{n}=n a$. In
the coupled-mode approach, we expand the electric field
$E(x,z)$ as a superposition of the fundamental modes centered
at each waveguide, $E(x,z) = \sum_{n} C_{n}(z) \psi(x - n a)$
where $\psi(x)$ is the waveguide mode. After posing  
$C_{n}(z) = C_{n} \exp(i \lambda z)$ and after inserting this into the paraxial wave
equation, one obtains the stationary equations for the
mode amplitudes
\be
( -\lambda + \epsilon_{n}) C_{n} + \sum_{m\neq n} V_{n m} C_{m} = 0,\label{eq:1}
\ee
where $V_{n m}$ is the coupling between the $n$th and $m$th waveguides. From Eq.(\ref{eq:1}), it is possible to formally express
\be
\epsilon_{n} = \lambda - \sum_{m\neq n} V_{n m} (C_{m}/C_{n}).\label{eq:2} 
\ee

Now, for a homogeneous system, we have $\epsilon_{n}=0$ and $C_{n}$ and $\lambda$ are computed by solving the simpler eigenvalue equation. For instance, for an infinite (and also semi-infinite) lattice with interactions to first nearest neighbors only, $C_{n}\equiv \phi_{n} \sim \sin(k n)$ and $\lambda = 2 V \cos(k)$, while for the case of an infinite lattice with first-and second nearest neighbors, $\phi_{n}\sim \sin(k n)$ and $\lambda = 2 V_{1} \cos(k) + 2 V_{2} \cos(2 k)$. For the case of a simultaneous presence of long range interactions and boundaries, a simple closed form for $\phi_{n}$ is not readily available, so is better to resort to a numerical $\phi_{n}$. Hereafter, we will keep the general notation $\phi_{n}$ for the unmodulated wavefunction. 

Now, in the spirit of Wigner and von Neumann\cite{wigner} we select an eigenstate $\phi_{n}$ of the homogeneous chain and proceed to modulate its envelope in such a way that the wave thus modulated is an eigenstate of the inhomogeneous system with eigenvalue $\lambda$:
\be 
C_{n} = \phi_{n} f_{n}
\ee
where $f_{n}$ is a decaying and normalizable envelope: $f_{n}\rightarrow 0$ as $n\rightarrow \infty$ and $\sum_{n} |f_{n}|^2 < \infty$. After inserting this ansatz into Eq.(\ref{eq:2}), we obtain
\be
\epsilon_{n} = \lambda - \sum_{m\neq n} V_{n m} \left(f_{m}\over{f_{n}}\right)\left( \phi_{m}\over{\phi_{n}} \right).\label{eq:4} 
\ee

\section{Results}
Let us now implement these ideas to four simple systems. We begin with the case of an infinite lattice with interactions to first neighbors only. Equation (\ref{eq:4}) reads
\be
\epsilon_{n} = \lambda - V \left(f_{n+1}\over{f_{n}}\right)\left( \phi_{n+1}\over{\phi_{n}} \right) - V \left(f_{n-1}\over{f_{n}}\right)\left( \phi_{n-1}\over{\phi_{n}} \right) .\label{eq:5} 
\ee

We take a monotonically decreasing envelope $f_{n}$ around some site $n_{0}$: 
\be 
\left( f_{n+1}\over{ f_{n}} \right) = (1 - \delta_{n})
\ee
where $\delta_{n}<1$. From this, we can solve formally for fn:
\be
f_{n} = \prod_{m=1}^{|n - n_{0}|-1} (1-\delta_{m})\label{eq:fn}
\ee
which can be rewritten as
\be f_{n} = \exp\left\{ \sum_{m=1}^{|n - n_{0}|-1}\log(1-\delta_{m})\right\}.
\ee
In the limit $n\rightarrow \infty$, and using that $\delta_{m}<1$, we can approximate
this by
\be
f_{\infty} \approx \exp\left\{ -\sum_{m=1}^{\infty}\delta_{m}\right\}
\ee
where we want $f_{\infty}\rightarrow 0$. This implies $\sum_{m=1}^{\infty} \delta_{m}=\infty$.

Now, we must choose an appropriate form for $\delta_{n}$ such that the local and decreasing ``potential'' $\{\epsilon_{n}\}$ remains bounded. From Eq.(\ref{eq:5}) we see that possible problems will emerge at the zeroes of $\phi_{n}$. To avert that, we choose
$\delta_{n}$ in the form
\be
\delta_{n} = {a\over{1 + |n|^{b}}} N^{2} \phi_{n+n_{0}}^{2} \phi_{n+n_{0}+1}^{2}\label{eq:10}
\ee
where $n\geq 1$ and $N$ is the length of the chain, $\sum_{n} |\phi_{n}|^2=1$, and the $f_{n}$ thus produced is symmetric around $n=n_{0}$. 
The reader can verify that, with the choice (\ref{eq:10}) for $\delta_{n}$, Eq.(\ref{eq:5}) can be written as
\begin{eqnarray}
\epsilon_{n} &=& \lambda - V (1-\delta_{|n+1-n_{0}|-1})\left( \phi_{n+1}\over{\phi_{n}} \right)\nonumber\\
 & &- V \left(1\over{1-\delta_{|n-n_{0}|-1}}\right)\left( \phi_{n-1}\over{\phi_{n}} \right) .\label{eq:new5} 
\end{eqnarray}
Therefore, when $\phi_{n}\rightarrow 0$, we have $\delta_{n}\rightarrow 0$, with the result that $\epsilon_{n}\rightarrow 0$ at those sites.
 
The particular choice of parameters $a$ and $b$ will determine the decay rate of both, the wavefunction envelope and the potential. On one hand, for practical purposes we need a potential that decays relatively fast, but we also need this potential not to be so abrupt that it can be indistinguishable from an impurity potential, since in that case our candidate state could be pushed out of the band. Using Eq.(\ref{eq:10}), The asymptotic decay
of the envelope at large $n$ values can be estimated in closed form, using the
Euler-Maclaurin formula, to be
\be
f_{n} \sim \exp(-\alpha_{n} |n-n_{0}|^{1-b}), \hspace{1 cm} n\rightarrow \infty\label{eq:12}
\ee
\begin{figure}[t]
\begin{center}
\noindent
\includegraphics[scale=0.425]{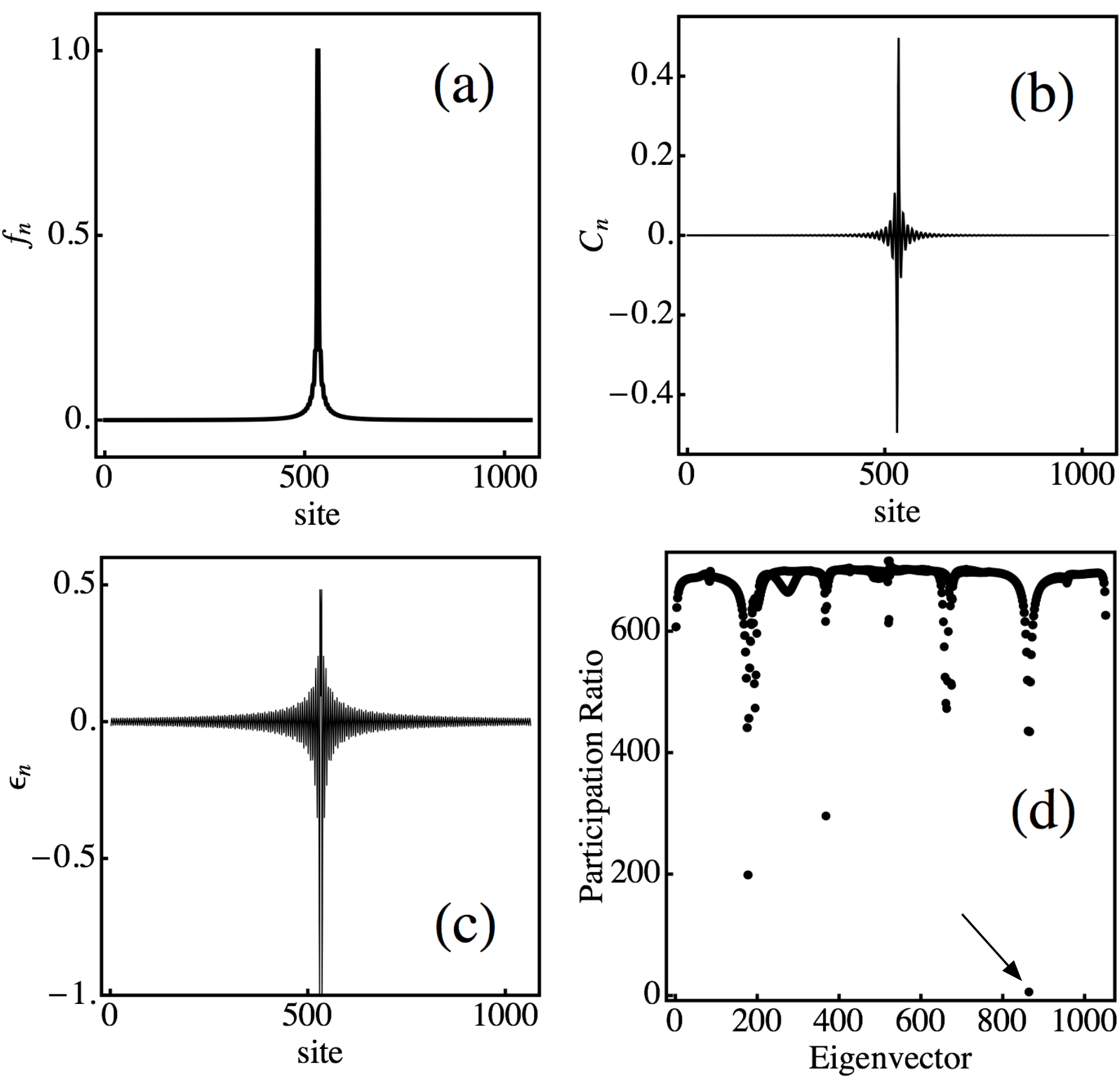}
\caption{Bulk BIC:(a) Discrete envelope function $f_{n}$ vs $n$. (b) Embedded mode profile.
(c) site energy distribution. (d) Participation ratio $R$ of all eigenvectors. The arrow indicates   the position of our BIC.
($N=1065, n_{0}=533, \lambda=1.56524, a=1/2, b=3/4$).}
\label{fig1}
\end{center}
\end{figure}
where $\alpha_{n}= B_{n} a /(1-b)$ with $B_{n}$ an oscillating factor of order $O(1)$ on average. Thus, for $0<b<1$ Eq.(\ref{eq:12}) gives a faster decay rate than the power law obtained by Wigner and von Neumann\cite{wigner}. On the other hand, based on Eq.(\ref{eq:10}), we can obtain the asymptotic behavior of the potential (Eq.(\ref{eq:5})) to be: 
\be
\epsilon_{n} \sim {A_{n} V a\over{|n-n_{0}|^{b}}},\hspace{1 cm} n\rightarrow \infty\label{eq:13}
\ee 
where $A_{n}$ is an oscillating factor of order $O(1)$ on average. Therefore, the envelope of the oscillating potential decays in space as a power law, qualitatively similar to the Wigner-Von Neumann case. From Eqs.(\ref{eq:12}) and (\ref{eq:13}) we see that if we make our BIC more localized by choosing a small value for $b$, then the site energy distribution will decay more slowly, and viceversa. 
A useful parameter to quantify the degree of localization of a state, is its participation ratio $R$, defined by, $R\equiv (\sum_{n} |C_{n}|^2)^2/\sum_{n} |C_{n}|^2$. For localized modes, $R\approx 1$ while
for extended states $R\approx N$, where N is the number of sites in the lattice. We will use $R$ as a measure of localization of all eigenstates in the presence of the $\{\epsilon_{n}\}$. In general we would like to have only one, or a few BICs generated by our procedure, keeping the rest as extended modes. This is yet another condition to be imposed on our selection of $a, b$. As we will see, these requirements can be met more easily when the system has only interactions to first nearest neighbors, rather than longer range couplings.

Figure 1 shows results for a lattice of $N=1065$ sites, where we have chosen $\lambda=1.56524$ out of the linear spectrum and 
used the trial function (\ref{eq:10}) with $a=1/2$, $b=3/4$ and $n_{0}=532$. While the BIC mode decays in space rather quickly ($\sim \exp(-\alpha\ n^{1/4})$), the site energy distribution decays at a slower pace ($n^{-3/4}$). Now, once in the possession of the distribution $\{\epsilon_{n}\}$, we proceed to find all the eigenvectors of the modulated array, and plot their participation ratio $R$, also shown in Fig.1. Our BIC is the one with the smallest $R$, namely $R\approx 6$ with the next higher one being $R\sim 200$. This means that we have a single BIC while the rest of the states are extended. 
Figure 2 shows the states inside the band that are closest in energy to the embedded state. The embedded state is the only state inside the band whose amplitude decreases to zero at large distance from the mode center, while all the rest of the band states are
extended. The presence of the modulation distorts the band states a bit but does not change their extended character. Examination of all eigenvalues reveals that a small number of them ($1\%$) fall outside of the band $[-2 V, 2V]$, and become impurity-like states. They were created as a consequence of the relative sharpness of the modulating potential chosen. An examination of the density of states (not shown) of the modulated system shows little change from the unmodulated case, save for the presence of some few states outside the band. This means our BIC is not inside some minigap, but is indeed inserted in a single band. Another good case is obtained for $a=1/2, b=0.9$ which gives rise to well-defined single BIC, plus nine modes outside the band. For some values of $a$, $b$ it is possible to have more than one BIC (e.g., for $n=1065$, $a=1/2$, $b=0.25$). In these cases, the number of impurity-like states outside the band is also observed to increase.

Let us now consider the case of a surface BIC, i.e., one where the BIC is centered at one of the ends of the lattice, say at $n=1$.  This case was recently considered by us\cite{molina}, but here we give a more general treatment. The general approach is the same as in the bulk case, but with an envelope that decreases away from the surface,
\begin{figure}[t]
\begin{center}
\noindent
\includegraphics[scale=0.4]{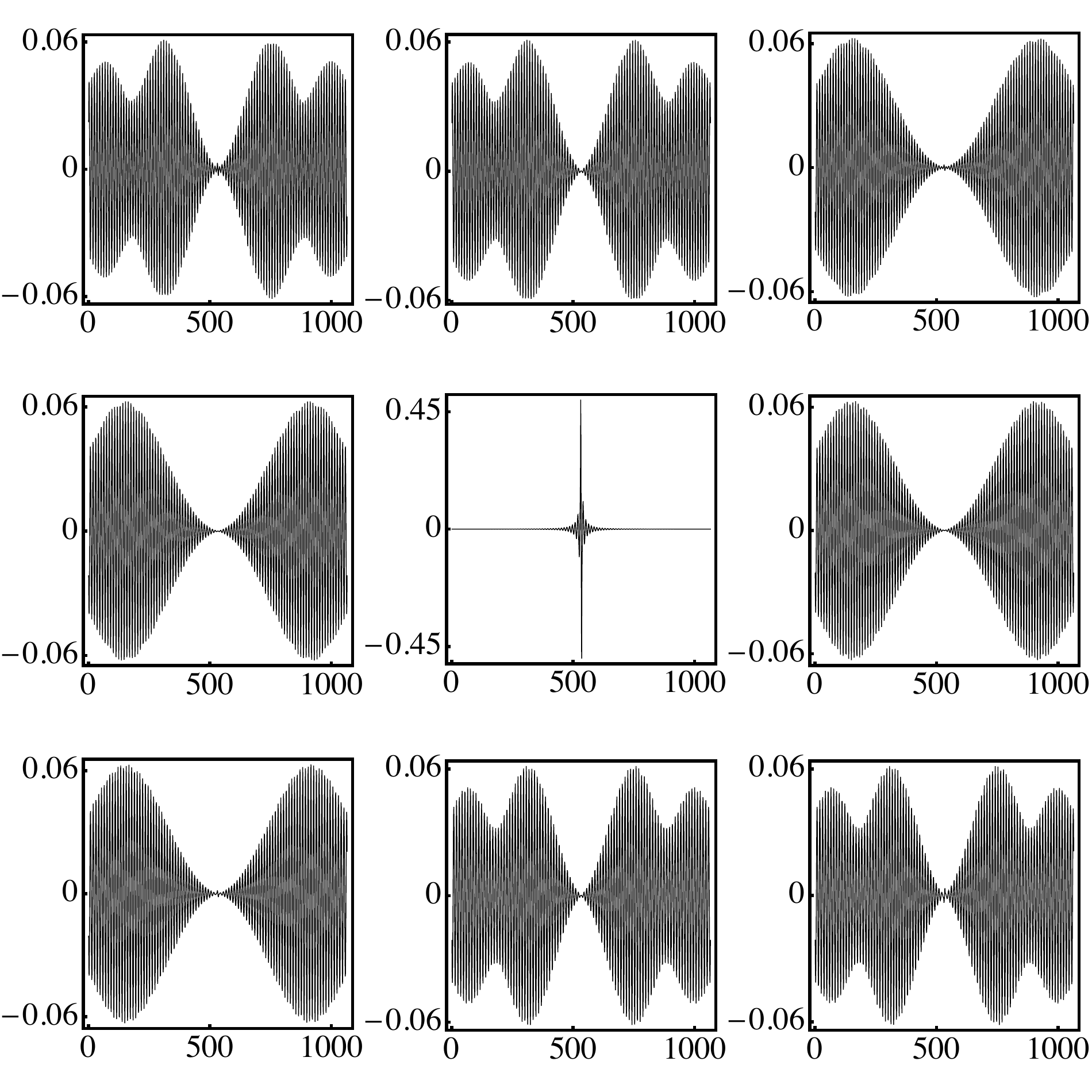}
\caption{States in the spectrum band that are the closest in
energy to the embedded mode (middle panel) ($N=1065, n_{0}=533, \lambda=1.56524, a=1/2, b=3/4$).}
\label{fig2}
\end{center}
\end{figure}
\be
f_{n} = \prod_{m=1}^{n-1}(1-\delta_{m})\label{eq:14}
\ee
for $n>1$, while $f_{1}=1$, and where we choose 
\be
\delta_{n}={a\over{n^{b}}} N^2 \phi_{n}^2 \phi_{n+1}^2.
\ee
\begin{figure}[t]
\begin{center}
\noindent
\includegraphics[scale=0.425]{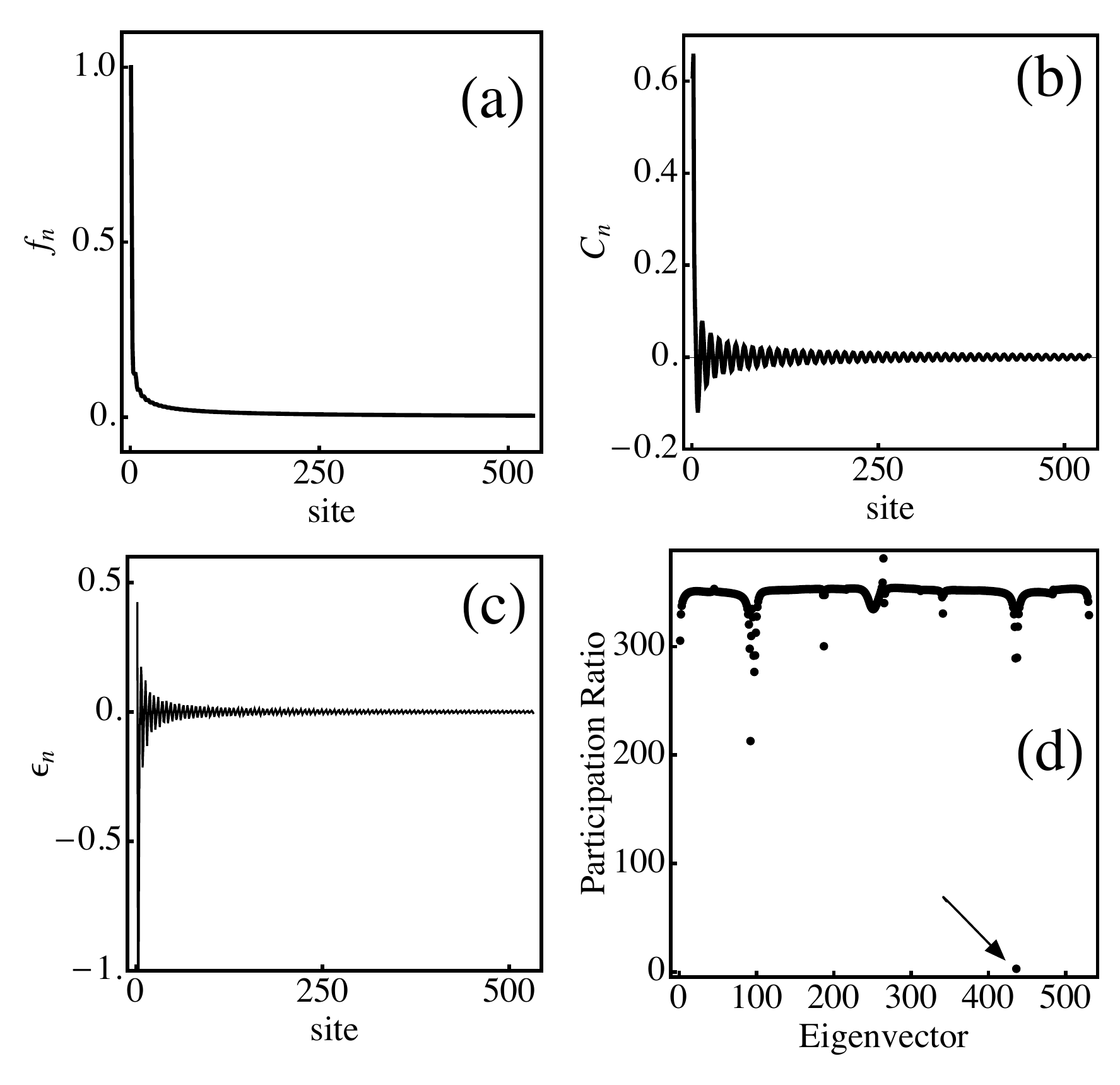}
\caption{Surface BIC: (a) Discrete envelope function $f_{n}$ vs $n$. (b) Embedded mode profile.
(c) site energy distribution. (d) Participation ratio $R$ of all eigenvectors. The arrow indicates   the position of our BIC. ($N=533, \lambda=1.69568, a=2/5, b=9/10$).}
\label{fig3}
\end{center}
\end{figure}
The site energy distribution is now given by
\be
\epsilon_{1}=\lambda-V\left({f_{2}\over{f_{2}}}\right)\left({\phi_{2}\over{\phi_{1}}}\right)
\ee
at the surface, and
\be
\epsilon_{n}=\lambda-V\left({f_{n+1}\over{f_{n}}}\right)\left({\phi_{n+1}\over{\phi_{n}}}\right)-
V\left({f_{n-1}\over{f_{n}}}\right)\left({\phi_{n-1}\over{\phi_{n}}}\right)\ \ \ \ \ \ 
\ee
for $n>1$. In this case we know that $\lambda=2 V \cos(k)$ and $\phi_{n}\sim \sin(k n)$, but we will keep the general notation. 
As in the previous case, different choices of $a$ and $b$ lead to different decay rates for the wavefunction $C_{n}=\phi_{n} f_{n}$ and its associated local potential $\epsilon_{n}$. Figure 3 shows a good trial wavefunction obtained with $a=2/5, b=9/10$. As in the bulk case, the local potential distorts most of the band modes, without altering their extended character. A small percentage of the initial modes get pushed outside the band, becoming impurity-like states, whose spatial extent depends on their proximity to the band. In the case of Fig.3, we have a single BIC, with $R\sim 3$ while the next higher $R$ is about $213$ and  only $3$ of the modes were pushed out of the band. Another good case is obtained for $a=0.5, b=0.9$ where we also have a single BIC, and there are only four states outside the band. It is also possible to have more than one BIC. For instance, for $a=1/3, b=1/2$ we have two BICs of similar participation ratios $R$. In this case, there are thirteen states outside the band. These features have also been observed in a recent work on surface BICs, using a modulation of the couplings\cite{longhi2013}. Sometimes, the extra BICs thus produced are not quite localized, but consist on a well-localized central peak surrounded by a very small but delocalized tail. They constitute resonance-like states rather than BICs. This is the case observed for $a=1/3, b=1/4$. The number of states pushed out of the band could in principle be reduced by taking a smoother $\epsilon_{n}$; the price to pay would be a local potential with a more extended range.

Let us now come back to the bulk BIC, but this time including interactions to first-and second nearest neighbors. The site energy distribution is now given by
\begin{eqnarray}
\epsilon_{n} &=& \lambda - V_{1} \left(f_{n+1}\over{f_{n}}\right)\left( \phi_{n+1}\over{\phi_{n}} \right) - V_{1} \left(f_{n-1}\over{f_{n}}\right)\left( \phi_{n-1}\over{\phi_{n}} \right)\nonumber\\
& & - V_{2} \left(f_{n+2}\over{f_{n}}\right)\left( \phi_{n+2}\over{\phi_{n}} \right) - V_{2} \left(f_{n-2}\over{f_{n}}\right)\left( \phi_{n-2}\over{\phi_{n}} \right)\ \ \ \ \ \ \ \label{eq:15}
\end{eqnarray}
where $V_{1}$ and $V_{2}$ denote the couplings to first and second nearest neighbors, respectively. 

The envelope $f_{n}$ is still given by Eq.(\ref{eq:fn}), but to avoid possible divergences and to keep $\epsilon_{n}$ bounded, we now choose $\delta_{n}$ in the form
\be
\delta_{n}={a\over{1 + |n|^b}} N^4 \phi_{n+n_{0}-1}^2 \phi_{n+n_{0}}^2 \phi_{n+n_{0}+1}^2 \phi_{n+n_{0}+2}^2 \label{eq:17}
\ee
where as before, $\phi_{n}$ is the (normalized to unity) extended mode we want to modulate (with eigenvalue $\lambda$) and $n_{0}$ is the position of the center of the localized state. Now, in spite of the more complex form of $\delta_{n}$, the asymptotic form for $C_{n}$ and $\epsilon_{n}$ is the same as in the case of coupling to nearest-neighbors only. This is due to the extended character of $\phi_{n}$ which does not affect the form of the decaying envelope of $\delta_{n}$ in Eq.(\ref{eq:17}), but only produces oscillations. 
Results are shown in Fig.4 for $N=1065, \lambda=1.69357, a=3.5, b=0.99$. In this example we take $V_{1}=1=V_{2}$ which corresponds to a geometrical arrangement where the guides are forming a `zigzag' structure. The presence of the extra coupling seems to increase the localization length of the BIC as compared to the case with only nearest neighbor coupling. This seems natural since an extra coupling would favor an increase of the spatial extent of the wavefunction.  In this case we also seem to have several BICs, although only the one we built  
\begin{figure}[t]
\begin{center}
\noindent
\includegraphics[scale=0.425]{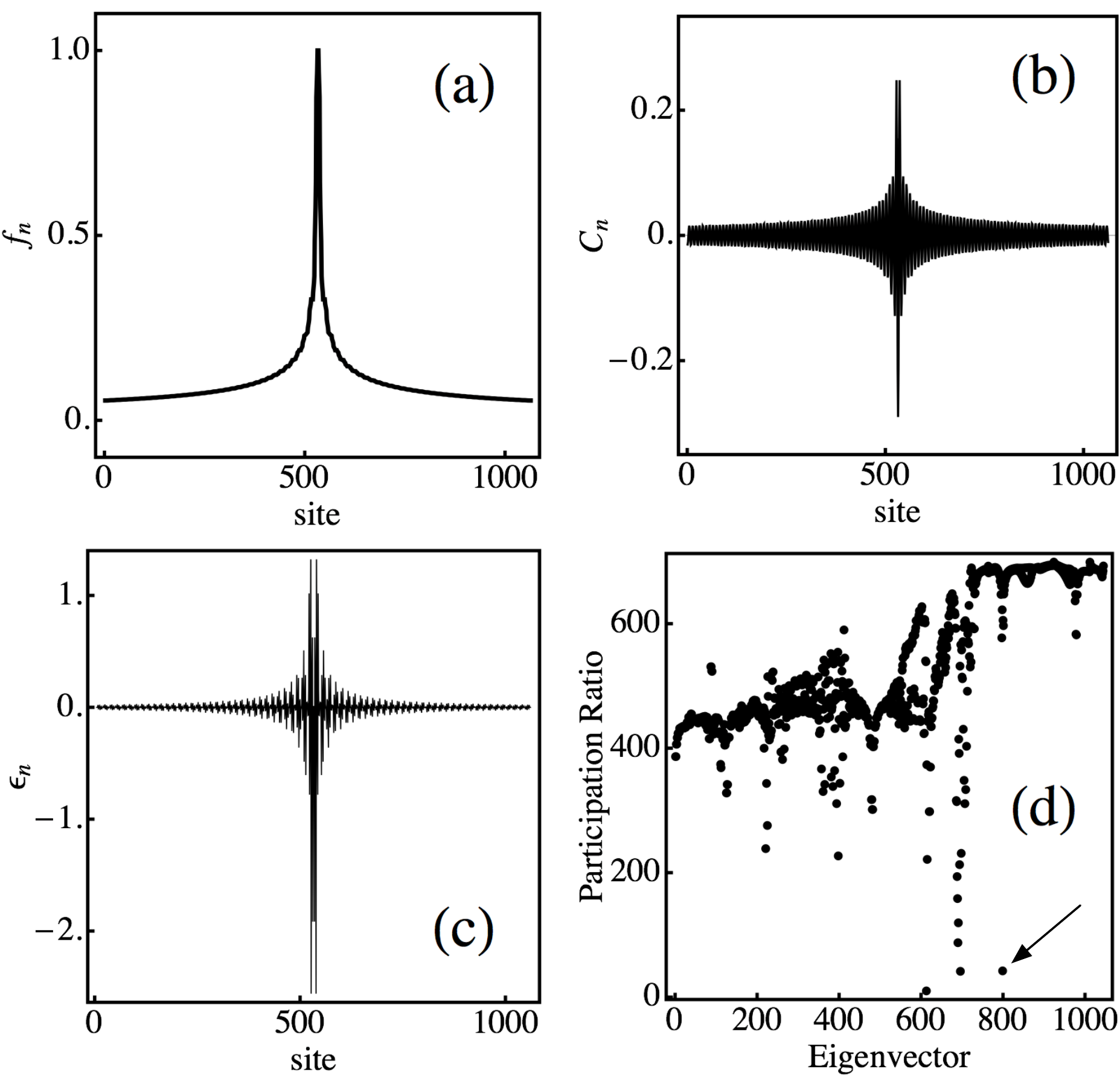}
\caption{Bulk BIC with long-range interaction: (a) Discrete envelope function $f_{n}$ vs $n$. (b) Embedded mode profile.
(c) site energy distribution. (d) Participation ratio $R$ of all eigenvectors. The arrow indicates   the position of our BIC.($N=1065, V_{1}=V_{2}=1, \lambda=1.69357, a=3.5, b=0.99$).}
\label{fig4}
\end{center}
\end{figure}
(eigenvector $799$) has a well-defined decaying rate (slow though); the other ones consist of  well-localized centers with non-decaying tails and seem to constitute resonances rather than {\em bona fide} BICs. Also, as a result of the potential modulation, about 20 states were pushed outside the band. We attribute this relatively high number to the steep shape of $\epsilon_{n}$ which can be visualized as a high impurity-like potential surrounded by a much smoother potential (see Fig.4c). The steep part is the one giving rise to the impurity-like states outside the band.

Finally, let us consider the case of a surface BIC in the presence of first-and second nearest-neighbor interactions. In this case it is not possible to obtain a simple, closed-form expression for the unmodulated eigenfunction $\phi_{n}$. The presence of interactions longer than nearest-neighbors complicates the use of the  image method technique\cite{long range}. Thus, $\phi_{n}$ and $\lambda$ have to be extracted directly from the initial diagonalization. The formal expression for the site energy distribution is
\begin{eqnarray}
\epsilon_{1}&=&\lambda - V_{1}\left({f_{2}\over{f_{1}}}\right)\left({\phi_{2}\over{\phi_{1}}}\right) - V_{2}\left({f_{3}\over{f_{1}}}\right)\left({\phi_{3}\over{\phi_{1}}}\right)\nonumber\\
\epsilon_{2}&=&\lambda -V_{1}\left( \left({f_{3}\over{f_{2}}}\right)\left({\phi_{3}\over{\phi_{2}}}\right) + \left({f_{1}\over{f_{2}}}\right)\left({\phi_{1}\over{\phi_{2}}}\right)  \right)\nonumber\\ 
&-& V_{2}\left({f_{4}\over{f_{2}}}\right)\left({\phi_{4}\over{\phi_{2}}}\right)\nonumber\\
\epsilon_{n}&=&\lambda 
- V_{1}\left( \left({f_{n+1}\over{f_{n}}}\right)\left({\phi_{n+1}\over{\phi_{n}}}\right) +
\left({f_{n-1}\over{f_{n}}}\right)\left({\phi_{n-1}\over{\phi_{n}}}\right)
  \right)\nonumber\\ 
&-& V_{2}\left(  \left({f_{n+2}\over{f_{n}}}\right)\left({\phi_{n+2}\over{\phi_{n}}}\right) +
\left({f_{n-2}\over{f_{n}}}\right)\left({\phi_{n-2}\over{\phi_{n}}}\right)
\right)
\end{eqnarray}
for $n>2$. 

Now, in order to avoid possible divergences at or near the zeroes of $\phi_{n}$, we choose the basic decay function $\delta_{n}$ in the form
\be
\delta_{n} = {a\over{n^b}} N^4 \phi_{n-1}^2\phi_{n}^2\phi_{n+1}^2\phi_{n+2}^2
\ee
for $n>2$, where we define $\delta_{1}=0$. With this choice, $\epsilon_{n}\rightarrow 0$ at those sites where $\phi_{n}\rightarrow 0$. The envelope function is then defined as in Eq.(\ref{eq:14}).

Figure 5 shows results for the case $N=533, V_{1}=1, V_{2}=0.5, \lambda=1.5655, a=5/2, b=3/4$. We notice that the site potential is more `bristling' than in the case of nearest-neighbors only. This lack of smoothness plus its relatively high value at the surface, causes that more states are now pushed outside the band ($30$ in the case of Fig.5). From Fig.5d we notice about five states that could correspond to BICs. A close examination of them reveals that only three of them (eigenvectors $252, 224$ and $381$) constitute true BICs while the other two (eigenvectors $71, 101$ and $102$) are resonances. The rest of the modes with larger $R$ are extended modes.

\begin{figure}[t]
\begin{center}
\noindent
\includegraphics[scale=0.455]{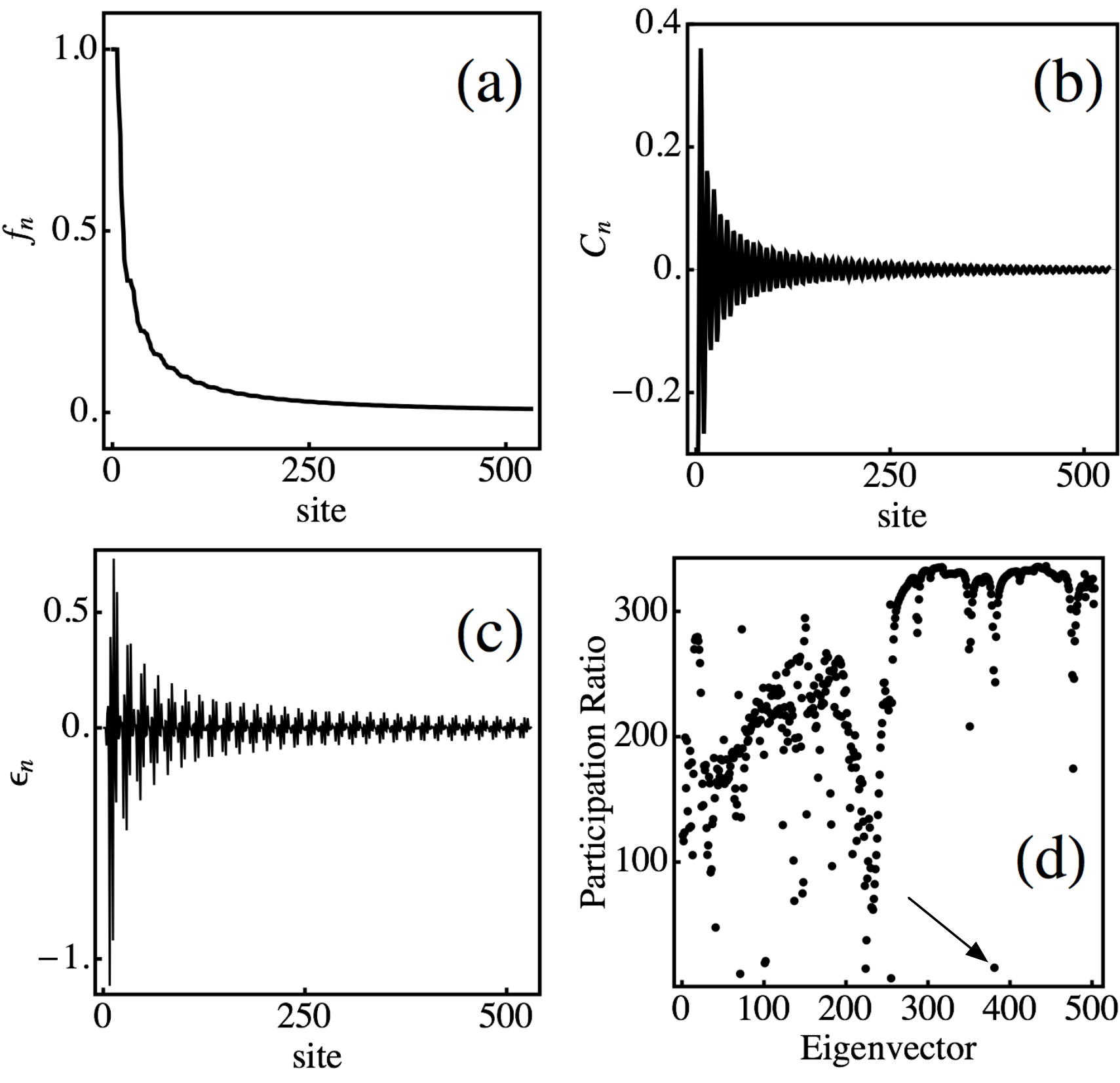}
\caption{Surface BIC with long-range interaction: (a) Discrete envelope function $f_{n}$ vs $n$. (b) Embedded mode profile.
(c) site energy distribution. (d) Participation ratio $R$ of all eigenvectors. The arrow indicates   the position of our BIC.($N=533, V_{1}=1, V_{2}=0.5, \lambda=1.5655, a=2.5, b=0.75$).}
\label{fig5}
\end{center}
\end{figure}

Finally, we performed a numerical structural stability analysis of each modulated system by adding some random noise to the site energy distribution: $\epsilon_{n}\rightarrow (1+\xi_{n})\epsilon_{n}$, where $\xi_{n} \in [-0.05,0.05]$ is a random quantity. In all cases examined it was observed that the perturbation did not affected the BIC appreciably, although it had an effect on the rest of the modes, when the BIC was a bulk mode. In this case, the rest of the modes  showed a general tendency towards localization, as evidenced by an overall decrease of their participation ratio (not shown). However, when the BIC was a surface mode, only minimal effects were observed. Of course, in all cases, the position of the BICs in the band was slightly shifted, depending on the random realization. The presence of absence of longer range interactions did not seem to play a significant role. We conclude that the BIC is structurally stable against small perturbations. This stability could be important when one tries to carry out an experimental realization of the systems described here; small errors in the building of the local potential $\epsilon_{n}$ are unavoidable and thus, the stability of the system give us hope that an experimental realization (perhaps in optics) can be soon carried on. 

\section{Conclusions}
We have examined bound states whose eigenenergies are embedded in the
continuous spectrum of a one-dimensional periodic lattice. We have demonstrated an explicit procedure to generate square-integrable, bulk and surface localized modes embedded in the continuum, considering interactions to first and second nearest neighbors. We have also given a prescription to generate the local bounded potential that gives rise to such modes. The envelope of these modes can be chosen to decrease in space faster than a power law, although the local potential decreases as a power law. The procedure of BIC generation can give rise to other BICs, as well as to resonances, and impurity modes located outside the continuous band. The BICs thus constructed are structurally stable against small perturbations.

\section{Acknowledgments}
This work was supported in part by Fondo Nacional de Ciencia y Tecnolog\'{\i}a (Grant 1120123), Programa Iniciativa Cient\'{\i}fica Milenio (Grant P10-030-F), and Programa de Financiamiento Basal (Grants FB0824 and FB0807).

\end{document}